\documentclass[letterpaper,11pt]{article}

\usepackage{achemso}
\usepackage{epsfig}

\begin{document}

\title{Magnetic Boron Nitride Nanoribbons with Tunable Electronic Properties }
\author{Veronica Barone and  Juan E. Peralta \\
\em Department of Physics, Central Michigan University, Mt. Pleasant, MI 48859}

\date{\today}

\maketitle

\begin{abstract}

We present theoretical evidence, based on total-energy first-principles calculations,
of the existence of spin-polarized states well localized at and extended along the edges of bare zigzag
boron nitride nanoribbons. Our calculations predict that all the magnetic configurations 
studied in this work are thermally accessible
at room temperature and present an energy gap. 
In particular, we show that the high spin state, with a magnetic moment of 1~$\mu_B$ at each edge atom, presents a 
rich spectrum of electronic behaviors as it can be controlled by applying an external electric field in order
to obtain metallic~$\leftrightarrow$~semiconducting~$\leftrightarrow$~half-metallic transitions.

\end{abstract}

\newpage

The recent experimental realization of atomically thin, long and narrow strips of graphene
(graphene nanoribbons, GNRs)\cite{geim2004,heer2006,kim2007} has sparked an intense research 
effort towards
the understanding of these novel materials  with promising applications 
in nanoelectronic and spintronic devices.
Albeit GNRs share many of the  properties of 
their close relatives, carbon nanotubes (CNTs),\cite{ribbon-dress,fujita1,fujita2}  graphene nanoribbons 
present reactive edges that dominate their electronic and magnetic behavior.\cite{kusakabe,yamashiro} 

The extreme importance of the edges in graphene has been 
pointed out by Wakabayashi et al.\cite{fujita1} by considering spin-polarized theoretical models:
While armchair 
nanotubes are all metallic, edge effects are so critical that bring zigzag ribbons into 
semiconductors with a spin-polarized ground state. This state is characterized by having 
opposite spins at each edge, coupled through two sublattices (bipartite lattice) spin-up and spin-down of the
hexagonal carbon network. As shown later by Son et al.,\cite{Son2006} this ground state can be potentially 
used as a spin-filter device.

Magnetism in low-dimensional systems involving $s$ and $p$ electrons like in graphene\cite{yaziev2007} 
and two-dimensional hexagonal boron nitride (h-BN)\cite{hBN-magnetism} 
is  still not entirely understood.
Two-dimensional h-BN is a large band 
gap insulator and hence, when rolled up into a tubular form, the so formed BN nanotubes 
remain large band gap insulators, despite their chirality and diameter.\cite{BNNT} 
It is then not surprising that nanoribbons 
made out of h-BN are also non-magnetic insulators, as reported recently by Nakamura et al. and Du et al., 
who theoretically studied the electronic properties of hydrogen-terminated 
boron nitride nanoribbons (BNNR).\cite{nakamura2005,du2007}
Yet, as edge effects in h-BN are expected to be as important as in GNRs,\cite{okada2000,okada2001} 
one can anticipate a different electronic behavior when considering spin-polarization in bare edged zigzag BNNRs.

In this work, we present theoretical evidence, based on total-energy first-principles calculations, 
of the existence of spin-polarized states well localized near and extended along the edges of zigzag BNNRs. 
We find that the electronic behavior of these magnetic materials can be controlled by external stimuli in order to switch between 
metallic~$\leftrightarrow$~semiconducting~$\leftrightarrow$~half-metallic electronic behavior.

All calculations have been performed within the density functional theory (DFT)  formalism using
periodic boundary conditions 
as implemented in the Gaussian Development Version program.\cite{gdv_short} 
We have employed the Heyd-Scuseria-Ernzerhof (HSE) hybrid functional,\cite{hse,hse-errata} which
incorporates a portion of non-local Hartree-Fock exchange in 
the short-range electron-electron interaction region.
Reported band gaps are calculated as HSE Kohn-Sham band energy differences, which  have been shown to provide
a very good approximation to electronic transitions
in  bulk solids,\cite{hse-bulk}  single walled CNTs,\cite{chirals,metallic}  and GNRs.\cite{Barone2006,Hod2007} 
More importantly for the purpose of this work, 
HSE provides the correct { \em qualitative} bang gap in small gap systems, where standard local and 
semilocal functionals fail.\cite{hse-bulk2,hse-bulk3}
The optimized geometries and the electronic structure of each of the systems under study in this work
have been obtained using the all-electron Gaussian basis set 3-21G, consisting of 5$p$ and 3$s$ combined into 3$s$ and 2$p$
Gaussian functions. Differences between band gaps obtained with this basis set and the larger 6-31G* basis set\cite{6-31g*}
are smaller than 0.05~eV, including geometry relaxation effects.
We have used 67 equally spaced $k$ points for the first Brillouin zone
integration (for a translational vector of 0.51~nm in the periodic direction).

We consider first the spin-compensated case of BNNRs of an approximate 
width of 1.6~nm with armchair (a-BNNR) 
and zizag (zz-BNNR) edges and both, bare and hydrogen terminated, as shown in Fig.~1.
It is not surprising to find that all hydrogen passivated BNNR as well as bare
a-BNNR are predicted to be large band gap insulators, as shown in Table~1
(results for the h-BN sheet are also shown for comparison.)\cite{Rubio-BN}
The fact that the band gap in these systems is smaller
than the one of the two-dimensional h-BN sheet has been already discussed by Nakamura et al.\cite{nakamura2005}
These authors found that due to edge effects, the band gap in passivated BNNR
decreases slightly with the ribbon width. Notably, we find that bare zz-BNNRs are 
metallic. These results are in line with the recent finding of Terrones et al.\cite{terrones2008} 
This behavior can be understood in terms of the electron delocalization 
imposed by the spin-compensated constraint. 
In view of these results we have performed spin-polarized
calculations in order to determine the magnetic nature of their ground state. 
Before going into more detail about the spin-polarized results, we would like to stress that this
behavior will only be found in bare zz-BNNRs, which 
can be potentially synthesized just as graphene nanoribbons,\cite{geim2004} and then forced into a heat 
treatment in order to clean the edges from chemisorbed functional groups. A similar experimental procedure has been already 
employed by Kobayashi et al. in graphene.\cite{kobayashi2005,kobayashi2006}

\begin{figure}[ht]
        \label{fig:1}
        \centerline{\epsfig{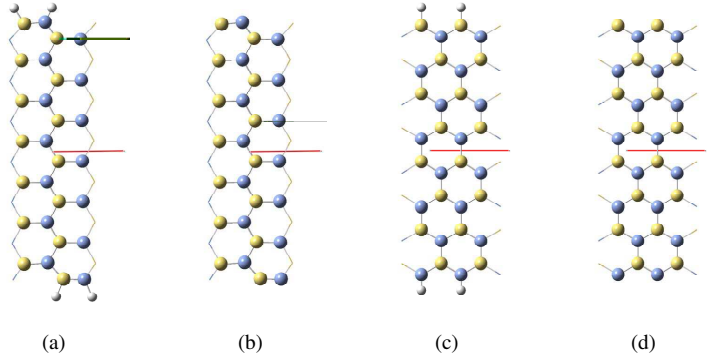}}
\caption{ Schemes of the four different models  under consideration: a) armchair passivated, 
b) armchair bare, c) zigzag passivated, and d) zigzag bare. B atoms appear in yellow 
and N atoms in blue. The translational vector is represented by a 
red line.}
\end{figure}

\begin{table}[ht]
   \caption{ \label{tb:table1}
   Comparison of energy band gaps (in eV) from spin-compensated calculations for the 
   systems shown in Fig.~1 and the 2-D hexagonal BN sheet.}
      \begin{center}
         \begin{tabular}{lc}
         \hline\hline
         Passivated a-BN                 &     5.82       \\[-0.2cm]
         Passivated zz-BN                &     5.56       \\[-0.2cm]
         Bare a-BN                       &     5.11       \\[-0.2cm]
         Bare zz-BN                      &     Metal      \\[-0.2cm]
         h-BN sheet                        &     5.92       \\[-0.2cm]
         h-BN sheet (GW approximation)\footnotemark[1]    &     6.00        \\
         \hline\hline
         \end{tabular} 
      \end{center}
      \footnote[1]~~Taken from Ref.~\citenum{Rubio-BN}. 

\end{table}

We will now turn to  the case of spin-polarized states in bare zz-BNNRs. 
We consider  a double periodicity unit cell (as shown in Figure 1-d) with
two B atoms at one edge and two N atoms at the opposite edge. This allows us to 
study the five simplest (domain wall) non-equivalent spin arrangements, as schematized in Fig.~2 
(in this figure we also introduce the notation for these five states.)
We find that all five spin-polarized configurations are energetically more favorable than the spin-compensated
solution. For instance, if we consider a 1.8~nm wide zz-BNNR, the energy difference between the
spin-compensated solution   and the least favorable magnetic configuration, \mbox{(++,+--)}, is as large as
560~meV/edge atom, evidencing that bare zz-BNNRs are magnetic materials.

\begin{figure}[ht]
        \label{fig:2}
        \centerline{\epsfig{figure=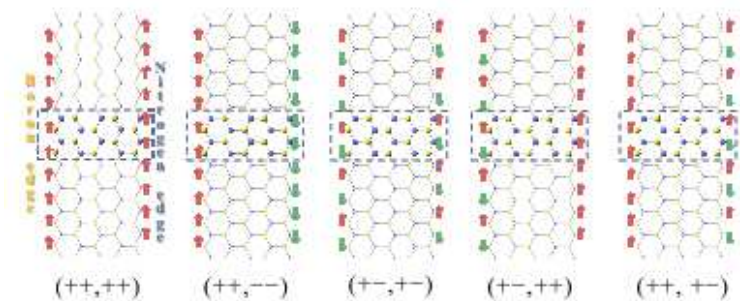, width=4.in, angle=0}}
\caption{Schematic representation of the five magnetic configurations studied for different
ribbon widths. Each state is denoted as a combination of two pairs of ``+'' or ``--'', according to the
magnetization on the B and N edges, respectively. }
\end{figure}

For all the zz-BNNRs studied in this work (with widths ranging from 0.9 to 1.8~nm), the ground
state solution corresponds to an antiferromagnetic spin arrangement at the B edge and a 
ferromagnetic arrangement at the N edge [corresponding to the \mbox{(+--,++)} magnetic configuration]
However,
all other magnetic solutions have comparable energies within 7~meV/edge
atom, indicating that all these states are thermally accessible at room temperature.

Mulliken spin population analysis assigns an atomic magnetic moment of about 1~$\mu_B$ 
for the  edge atoms (B and N), independently of the magnetic configuration, evidencing 
a strong localization of the spin density at
the edges. The energy difference of these two spin chains,  \mbox{E(++,++) $-$ E(++,-- --)}, shown in Table~2,
provides an estimate  of the inter-spin chain interaction. 
This energy difference rapidly vanishes as the ribbon
reaches a width of  $\sim$~1.4 nm. From that point on, the interaction between both spin chains becomes negligible
and the magnetization quenches through the inert boron nitride
hexagonal backbone. This behavior contrasts with  the one found
in graphene nanoribbons, where both zigzag edges communicate via an
antiferromagnetic spin arrangement mediated by the carbon backbone.
In Table~2 we also present the energy per edge atom relative to the ground state
for  the \mbox{(+--,+--)} and \mbox{(++,+--)}   spin configurations.
The energies of the two chains (B and N) are
additive: The energy necessary  to convert the \mbox{+-- B} chain into a \mbox{++ B} chain is about
6 meV/edge atom, while the energy required to convert the N spin chain from \mbox{++} to
\mbox{+--} is 0.4 meV/edge atom. As these chains become less interacting, the energy
required to access the \mbox{(++,+--)} solution from the ground state, \mbox{(+--,++)}, can be approximated as  the 
sum of the individual energies, about 6.4 meV/edge atom.

\begin{table}[ht]
 \caption{ \label{tb:table2}
Relative energies (with respect to the ground state) for different magnetic configurations and widths of zz-BNNR.}
        \begin{center}
           \begin{tabular}{ccccc}
           \hline\hline
    &    \multicolumn{4}{c}{Energy (meV/edge atom)}   \\
   \cline{2-5}
width (nm) & (++,++) & (++,-- --) & (+ --, + --)  &  (++, + --) \\
   \cline{1-5}
 0.93      & 5.0    & 4.8       & 0.4          & 5.3      \\[-0.2cm]
 1.15      & 5.3    & 5.7       & 0.4          & 5.7     \\[-0.2cm]
 1.37      & 5.6    & 5.6       & 0.4          & 6.0      \\[-0.2cm]
 1.59      & 5.8    & 5.8       & 0.4          & 6.2      \\[-0.2cm]
 1.81      & 6.0    & 6.0       & 0.4          & 6.4      \\
           \hline\hline
   \end{tabular}
 \end{center}
\end{table}

Further evidence of the localized character of these spin chains can be found in 
the density of states. As shown in Fig.~3, $p$ orbitals from the B edge
atoms are well localized in the valence region and do not hybridize with the orbitals from the BN
backbone. 
On the other hand, $p$ orbitals from edge N atoms
hybridize with the neighboring atoms from the BN network and dominate the conduction region,
but yet magnetic moments from the BN network on the N side also quench rapidly. These
results agree with the theoretical predictions in defective BN nanotubes of Kang,\cite{kang2006}
who finds that B edges present higher
stability of the local spin configuration than N edges. 

The HSE functional predicts all five spin configurations to be semiconducting,
 which contrasts with the metallic character of the energetically unfavorable
spin-compensated solution. We would like to note that according to our calculations, 
the semi-local functional PBE\cite{PBE_1,PBE_2} does not open a band gap for the minority spin channel, 
thus predicting a half-metallic behavior for bare zigzag BNNRs. 
A similar trend has been observed in the case of oxidized graphene nanoribbons for which 
PBE fails to predict the spin-polarized semiconducting behavior.\cite{Hod2007}

\begin{figure}[h]
        \label{fig:3}
         \centerline{\epsfig{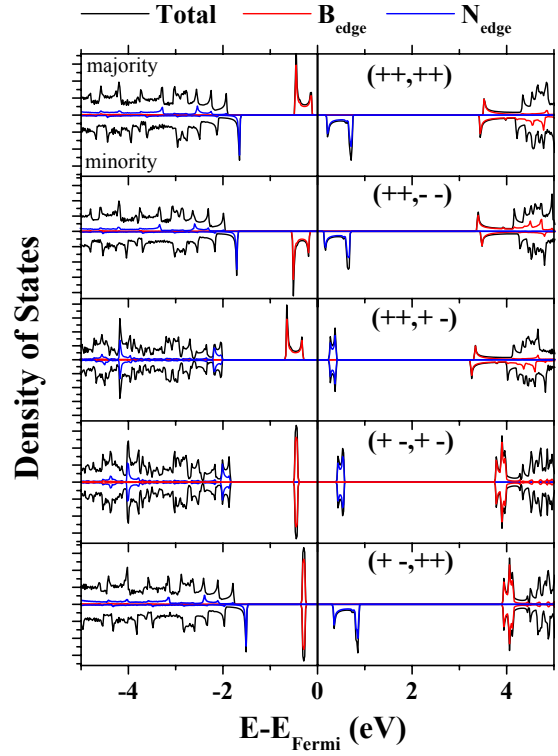}}
\caption{Total and partial density of states of the five spin configurations studied in this work
for a 1.59 nm wide bare zz-BNNR. In black we show the total density of states and in red and blue the 
partial density of states from the two 
B and the two N edge atoms, respectively.}
\end{figure}

As reported by Son et al.,\cite{Son2006} zigzag graphene nanoribbons undergo a transition from
semiconducting to half-metallic due to charge reorganization in the edges
if a strong enough electric field is applied in
the transverse direction.
Therefore, the question arises of how these systems with
strongly localized magnetization and more ionic nature than graphene 
behave in the presence of an external transverse electric field.
Contrary to the case of graphene, BN ribbons are not symmetric in the
transverse direction and therefore the effect of an electric field in
the B$\rightarrow$N and B$\leftarrow$N directions must be considered. In
Fig.~4  we show our results for the band gap as a function of the field intensity and direction for all five
magnetic configurations, i.e.:  \mbox{(+--,++)}, \mbox{(+--,+--)}, \mbox{(++,+--)}, \mbox{(++,-- --)}, and \mbox{(++,++)}.
The \mbox{(+--,++)}, \mbox{(++,+--)}, and \mbox{(+--,+--)} states undergo a transition from
semiconducting to fully metallic for different field strengths, depending on the
direction of the field. As the total magnetic moment of the unit cell in the
\mbox{(+--,+--)} case is zero, there is a symmetry in the behavior of the majority and minority
spin channels that is not present in the \mbox{(+--,++)} and \mbox{(++,+--)} cases that exhibit a spin
magnetic moment of 2~$\mu_B$ per unit cell. 
The \mbox{(++,++)} and \mbox{(++,-- --)} states 
present a richer spectrum of electronic
behaviors. 
The \mbox{(++,-- --)} configuration shows a large band gap of  5.36~eV for the majority spin channel and a small gap of
0.34 eV for the minority spin channel at zero field. When the field is such that electrons are reorganized 
towards the B edge, the system becomes metallic in both spin channels for a field of about 0.6~eV/\AA. However, 
when the field is applied in the opposite direction, the band gap of the minority spin channel closes while 
the band gap of the majority spin channel decreases only slightly, thus presenting a half-metallic behavior
for this particular direction of the electric field.
The \mbox{(++,++)} state presents a high magnetic moment of 4~$\mu_B$ per cell. 
When no external field is applied, this state is semiconducting, with
large band gaps of 3.64 eV and 1.86 eV for the majority and minority
spin channels, respectively. When an electric field in the B$\leftarrow$N
direction is applied, the electronic reorganization produces a rapid closing of
the band gap for both spin channels, transforming these materials into
metals. However, when the field is applied in the opposite direction (B$\rightarrow$N) we
do not observe any noticeable effect until the external field reaches an intensity
of 0.3~V/\AA. At this point, the band gap closes rapidly for both channels up to a
field of about 0.5~V/\AA at which the band gap of the minority spin channel closes
completely while the band gap of the other channel increases again.
Therefore, for this particular direction of the external field there is a range of
intensities for which the material
behaves as a half-metal. 
Our calculations indicate that, as in the case of H-terminated BN ribbons,\cite{nakamura2005}
the band gap of the minority spin channel slowly decreases as the width of the ribbon increases. This effect produces
the field strength required in order to achieve the half-metallic behavior   
also to slightly decrease with the ribbon width.

\begin{figure}[h]
        \label{fig:4}
        \centerline{\epsfig{figure=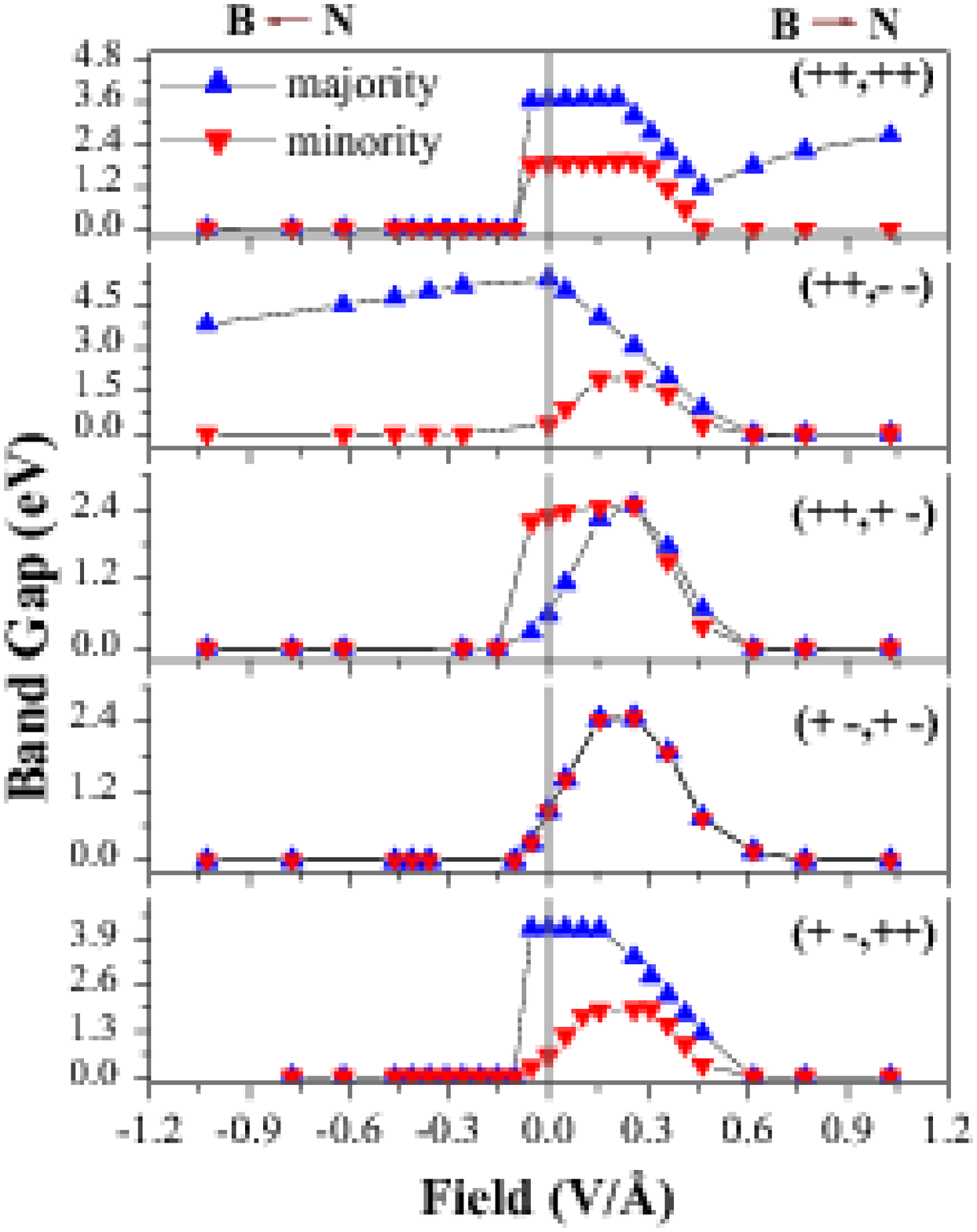, width=3.0in, angle=0}}
\caption{Band gaps as a function of the field intensity and direction 
for the five spin configurations studied in this work for a 1.59~nm wide bare zz-BNNR
(positive field corresponds to B$\rightarrow$N and negative field to B$\leftarrow$N.)}
\end{figure}

Even though both magnetic solutions \mbox{(++,++)} and \mbox{(++,-- --)} present this broad range of electronic behaviors,
it is the high spin state \mbox{(++,++)} the most interesting for potential device applications as it can be 
stabilized over the other magnetic configurations by applying an external magnetic field.
The interplay between the field strength and direction, and the electronic behavior of the \mbox{(++,++)} state
is also clearly manifested in the density of states.
In the right panel of Fig.~5 we show the effect of an electric field of different intensities 
applied in the B$\leftarrow$N 
direction and how rapidly both channels exhibit a finite DOS at the
Fermi level. 
Applying an electric field in the opposite direction (B$\rightarrow$N), as shown in the left panel of Fig.~5, 
increases the electron population at the B edges while decreasing it at the N edges.
As a result, occupied states from
the N edge shift towards the Fermi level leading to a single metallic channel.

\begin{figure}[h]
        \label{fig:5}
        \centerline{\epsfig{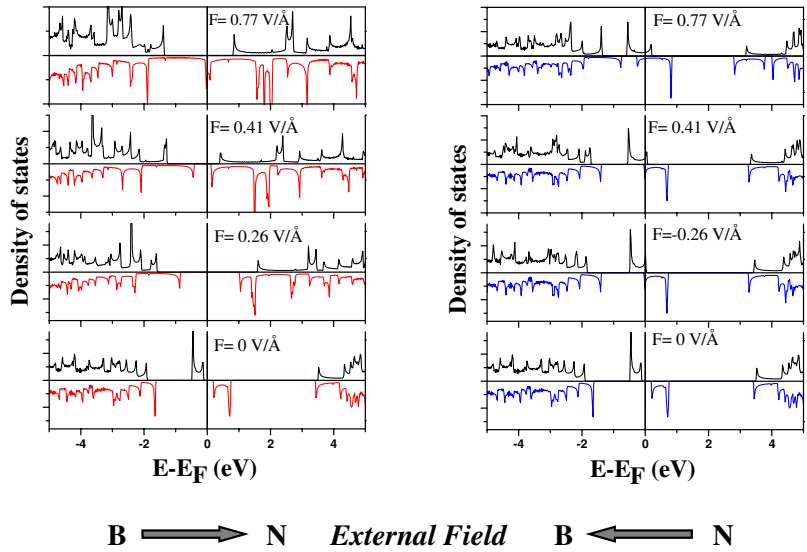}}
\caption{Total density of states of the \mbox{(++,++)} configuration for the
1.59~nm wide bare zz-BNNR and different electric field intensities. 
The field direction is indicated below each panel.}
\end{figure}

In summary, we have studied BNNRs with bare zigzag edges and found that these materials
are magnetic semiconductors 
with an energy gap that decreases slightly with the ribbon width.
Even though all spin configurations in zz-BNNRs are thermally accessible at room temperature,
as in the case of grephene nanoribbons, 
there is an important difference: the target state for practical applications
in the case of zz-BNNRs is the high spin state, \mbox{(++,++)}, while in graphene 
it is the \mbox{(++,-- --)} state. 
This high spin state can
be stabilized over the other magnetic configurations by applying an external magnetic field.
Then, an applied transverse electric field
will produce electron reorganization towards the B edge or N edge, depending on its
direction,  enabling an external control of the band gap of zz-BNNR to produce
metallic~$\leftrightarrow$~semiconducting~$\leftrightarrow$~half-metallic transitions.

{\Large \bf Acknowledgements}

J.E.P. acknowledges support from the President's Research Investment Fund (PRIF) 
and a start-up grant from Central Michigan University. Some of the calculations
were made possible due to the NSF/MRI grant PHY-0619407.


\providecommand{\url}[1]{\texttt{#1}}
\providecommand{\refin}[1]{\\ \textbf{Referenced in:} #1}

\end{document}